\documentstyle[prl, aps, amssymb
]{revtex}
\begin{document}
\draft
\wideabs{
\title{$^{63}$Cu NQR evidence of 
dimensional crossover to anisotropic 2d regime in $S=\frac{1}{2}$ 
three-leg ladder 
Sr$_{2}$Cu$_{3}$O$_{5}$}
\author{K. R. Thurber,$^{1,2}$ T. Imai,$^{1,2}$ T. Saitoh,$^{3}$ M. 
Azuma,$^{3}$ M. Takano,$^{3}$ and F.C. Chou$^{2}$}
\address{$^{1}$Department of Physics and $^{2}$Center for Materials Science and 
Engineering, M.I.T., Cambridge, MA 02139}
\address{$^{3}$Institute of Chemical Research, Kyoto University, Uji, Kyoto 611, 
Japan}
\date{\today}
\maketitle
\begin{abstract}
We probed spin-spin correlations  up to 725 K with $^{63}$Cu NQR
 in the $S=\frac{1}{2}$ three-leg ladder Sr$_{2}$Cu$_{3}$O$_{5}$.  
We present experimental evidence that below 300 K, weak inter-ladder coupling 
causes dimensional crossover of the spin-spin correlation length $\xi$ from quasi-1d 
($\xi\sim\frac{1}{T}$) to anisotropic 2d regime ($\xi\sim \exp\frac{2\pi\rho_{s}}{T}$, 
where $2\pi\rho_{s}=$290$\pm$30 K is the effective spin stiffness).  This is the first 
experimental verification of the renormalized classical behavior of 
the anisotropic non-linear sigma model in 2d, which has been recently proposed for the striped phase in high 
$T_{c}$ cuprates.
\end{abstract}
}
Quantum magnetism of spin $S=\frac{1}{2}$ in low dimensions has 
attracted strong attention 
since the discovery of 
high $T_{c}$ superconductivity in hole-doped 
CuO$_{2}$ planes.  Intensive experimental and 
theoretical studies have led to a reasonably good understanding of 
the statistical properties of the 2d square-lattice\cite{Kastner}.
More recently, the successful synthesis of SrCu$_{2}$O$_{3}$ and Sr$_{2}$Cu$_{3}$O$_{5}$ 
 with 
two- and three-leg ladder structure, respectively 
\cite{Azuma}, has made it possible to investigate $S=\frac{1}{2}$ quantum spins 
between 1d and 2d.  Subsequent discovery of 
superconductivity in another two-leg ladder material 
(La,Ca,Sr)$_{14}$Cu$_{24}$O$_{41}$ \cite{Uehara} further enhanced interest in 
the statistical properties of quantum spin ladders.  In addition, under-doped 
high $T_{c}$ superconductors show a stripe structure 
\cite{Tranquada,Allen} which has been proposed as undoped 
three-leg spin ladders coupled across rivers of doped holes.  
Concentrated theoretical effort is underway to reproduce this 
phenomenon based on the anisotropic non-linear sigma model 
\cite{anisotropic}.  This extends the fundamental 
importance of spin ladders beyond the field of magnetism to the 
mechanism of high $T_{c}$ superconductivity.   

Uniform susceptibility \cite{Azuma,Johnston}, 
NMR/NQR \cite{Ishida,Carretta,Magishi,Takigawa,Imai}, and neutron scattering measurements 
\cite{Eccelston,Azuma3} have confirmed the most basic prediction
 about spin ladders, namely that ladders with an even number of legs have a spin 
gap $\Delta$ in their 
spin excitation spectrum, while ladders with an odd number of legs do 
not \cite{Dagotto}.  For SrCu$_{2}$O$_{3}$ and Sr$_{2}$Cu$_{3}$O$_{5}$, spin susceptibility by 
Azuma, {\it et al.} \cite{Azuma} and pioneering high field $^{63}$Cu 
NMR between 100 K and 300 K by Ishida, {\it et al.} 
\cite{Ishida} revealed the qualitative difference of magnetic 
properties between two- and three-leg ladders in the low temperature 
limit.  However, the nature of spin correlations at finite temperatures
remains largely 
untested, despite recent theoretical developments \cite{Sachdev,Greven,Young-jun,gopalan,Sandvik}.  
The large exchange 
interaction $J=$ 1300$\sim$1600 K in copper-oxide ladders 
\cite{Azuma,Johnston,Imai,Eccelston} makes measurements of spin-correlations at 
elevated temperatures essential to understanding the statistical properties at finite 
temperatures.

In this Letter, we report the first $^{63}$Cu NQR measurements of the $^{63}$Cu nuclear 
spin-lattice relaxation rate $1/T_{1}$ \cite{T1NQRNMR} and the Gaussian component of 
spin-spin relaxation rate $1/T_{2G}$ \cite{T2Redfield} for the  
three-leg ladder Sr$_{2}$Cu$_{3}$O$_{5}$ from 83 K up to 725 K.  The Gaussian 
component of the $^{63}$Cu nuclear spin-spin relaxation rate, $1/T_{2G}$, probes the  
spin-spin correlation length, $\xi$, 
as demonstrated earlier for the 2d 
square-lattice\cite{Pennington,Imai2D,Thurber} 
and the 1d spin-chain\cite{Takigawa1D}.  
We present evidence that the  
spin-spin correlation length $\xi$ in the three-leg ladder follows the 
1d form $\xi\sim\frac{1}{T}$ above 300 K.  To our surprise, 
however, we found that weak inter-ladder coupling along the c axis results in dimensional 
crossover to a quasi-2d regime below 300 K, where $\xi$ diverges 
exponentially, $\xi\sim \exp\frac{2\pi\rho_{s}}{T}$ 
($2\pi\rho_{s}$=290$\pm$30 K is the 
effective spin stiffness).  
The anisotropic non-linear sigma model 
\cite{anisotropic} successfully describes the 
static and dynamic NQR/NMR properties in the quasi-2d regime.  To 
the best of our knowledge, this is the first experimental 
demonstration of the validity of the anisotropic non-linear 
sigma model in a quantum Heisenberg antiferromagnet.   

In Figure 2(a) and (b), we compare the temperature dependence of the 
nuclear spin relaxation rates for the three-leg ladder 
Sr$_{2}$Cu$_{3}$O$_{5}$, with other geometries between 1d and 2d:  1d spin-chain 
Sr$_{2}$CuO$_{3}$, two-leg ladders SrCu$_{2}$O$_{3}$ and 
La$_{6}$Ca$_{8}$Cu$_{24}$O$_{41}$, and 2d square-lattice 
Sr$_{2}$CuO$_{2}$Cl$_{2}$.  The polycrystalline samples of the 
ladders, Sr$_{2}$Cu$_{3}$O$_{5}$ and SrCu$_{2}$O$_{3}$, were grown 
at Kyoto under high pressure \cite{Azuma}, while the single crystal for 
1d spin-chain Sr$_{2}$CuO$_{3}$ was grown at M.I.T. using an optical floating 
zone furnace.  The results for 2d square-lattice 
Sr$_{2}$CuO$_{2}$Cl$_{2}$ are from \cite{Thurber} and for two-leg 
ladder La$_{6}$Ca$_{8}$Cu$_{24}$O$_{41}$ from \cite{Imai}.  

Technically, the present NQR approach used for Sr$_{2}$Cu$_{3}$O$_{5}$ 
and SrCu$_{2}$O$_{3}$ is much 
harder than the high field NMR method, 
because the low resonance frequency of NQR (11.32 MHz at 300 K for the 
edge chain in Sr$_{2}$Cu$_{3}$O$_{5}$) makes the signal intensity 
two orders of magnitude weaker.  
However, NQR allows us to conduct measurements at elevated temperatures 
limited only by sample decomposition.  In addition, the three-leg 
ladder structure shown in figure 1(a)\&(b) has two different copper 
sites, the atoms in the central chain and those in the two edge 
chains.  Our 
experiments measure 
the relaxation times for the two sites separately without 
line superposition.  Separation of the two signals is essential because 
$1/T_{1}$ differs by a factor of 2.3 \cite{rates}.   

$1/T_{1}$ measures the imaginary part of the dynamical electron spin 
susceptibility $\chi''({\bf q},\omega_{n})$ at the NQR/NMR frequency 
$\omega_{n} (\approx$ 11 (NQR) or 100 (NMR) MHz), or low-frequency spin 
dynamics \cite{Moriya}.
\begin{equation}
          \frac{1}{T_{1}} = k_{B}T\frac{ \gamma_{n}^{2}}{\mu_{B}^{2} \hbar} 
	\sum_{{\bf q}} | F({\bf q}) |^{2} \frac{\chi''({\bf 
	q},\omega_{n})}{\omega_{n}}
\label{T1}
\end{equation}
where $\gamma_{n}$ is the nuclear 
gyromagnetic ratio.  $F({\bf q})$ is the wave vector 
dependent hyperfine form factor\cite{Moriya}. 
On the other hand $1/T_{2G}$ measures the real part 
of the electron spin susceptibility at zero frequency, $\chi'({\bf q})$
\cite{Pennington}.
\begin{equation}
\left( \frac{1}{T_{2G}} \right)^{2} \approx \frac{0.69 
(\gamma_{n}\hbar)^{4}}{8\hbar^{2}} \sum_{q} | F({\bf q}) |^{4} 
\chi'({\bf q})^{2}
\end{equation}
where the factor 0.69 is the isotope abundance of $^{63}$Cu.  
$1/T_{2G}$ provides quantitative information of spin-spin correlation 
length $\xi$ in low dimensional systems; $1/T_{2G}\sim$$T\xi$ for 
2d square-lattice \cite{Imai}, $\sim\sqrt{\xi}/T$ for ladders at 
 elevated temperatures 
(see below), and $\sim 1/\sqrt{T}$ for 1d spin-chain \cite{Takigawa1D}.

The results in Fig.2 clearly show dramatic change of spin-spin 
correlations with dimension.  In 2d square-lattice Sr$_{2}$CuO$_{2}$Cl$_{2}$, 
$1/T_{1}\sim T^{3/2}\xi$ 
\cite{Chak-Orb} and $1/T_{2G}\sim T\xi$ \cite{Imai} diverge 
exponentially, because $\xi$ diverges exponentially in the 
{\it renormalized classical} low temperature regime \cite{CHN,Hasenfratz},  
\begin{equation}
\xi= \frac{e}{8}\frac{\hbar c}{2\pi\rho_{s}} \exp \left( 
\frac{2\pi\rho_{s}}{T} \right) \left[ 1 - \frac{T}{4\pi\rho_{s}} + 
O(T^{2}) \right].
\end{equation}
where $c$ is the spin wave 
velocity, and $2\pi\rho_{s}(=1.13J)$ is the spin stiffness.  
As shown in Fig.3(a), the semi-logarithm plot of $1/T_{1}T^{3/2}$ and 
$1/T_{2G}T$ shows a linear behavior with slope $2\pi\rho_{s}= 1700 \pm 
150$ K.  
In 1d spin-chain Sr$_{2}$CuO$_{3}$, $1/T_{1}$ and $1/T_{2G}$ show much 
weaker temperature dependence up to 800 K.  This agrees with earlier 
confirmation of theoretical predictions below 300 K by Takigawa et al.\cite{Takigawa1D}, 
$1/T_{1}\sim$ constant and $1/T_{2G} \propto 1/\sqrt{T}$ \cite{Starykh}.
In two-leg ladder SrCu$_{2}$O$_{3}$, the presence of spin-gap 
$\Delta\sim$ 450 K 
\cite{Azuma,Imai} in the spin excitation spectrum results 
in exponential decrease of $1/T_{1}$ \cite{Ishida,Imai}
below $\sim$ 450 K, where $1/T_{2G}$ (and thus $\xi$ 
\cite{Greven,Sandvik}) gradually saturates.  

The results for three-leg ladder 
fall between 2d square-lattice and 1d spin-chain, as expected.  
The mild temperature dependence above 300 K is easy to understand.  
According to recent weak coupling continuum theory by 
Buragohain and Sachdev\cite{Sachdev}, which is
applicable 25 K $\ll 
T \lesssim$ 500 K for $J=$ 1500 K, the  spin structure 
factor is given as $\chi(q) = S(q) / k_{B}T \propto (1/k_{B}T) 
(\ln(T/\Lambda_{MS}))^{2} \xi/(1 + 
(q-\pi)^{2}\xi^{2} )$ with $\Lambda_{MS}$ roughly predicted as 25 K.  Inserting this expression into 
eq.(2), the leading order temperature 
dependence of $T_{2G}$ is given as $T_{2G} \propto 
T/\sqrt{\xi}$, with logarithmic corrections that 
become significant at low temperatures.  Since the spin-spin correlation 
length in a three-leg ladder is given as $\xi \sim \frac{1}{T}$ 
\cite{Sachdev,Greven}, 
we expect $T_{2G} \sim T^{3/2}$.  Indeed, as shown in the inset to Fig. 
2(b), we found power law behavior with exponent $\frac{3}{2}$ above 
300 K.  Comparison of $1/T_{1}$ with the 
analytic theory is more difficult because we cannot estimate the large 
diffusive (i.e. $q=0$) contributions in the quasi-1d regime.  Further numerical calculations 
are necessary to understand the observed constant 
behavior of $1/T_{1}$ at $T\sim\frac{J}{2}$.

In contrast with the mild temperature dependence at elevated temperatures, the 
divergent behavior below 300 K is quite surprising.  As shown in 
Fig.3, {\it the temperature dependence is exponential, similar to the 
case of the 2d square-lattice.}  The linearity in the 
semi-logarithmic plot extends for an order of magnitude.  Within the 
framework of an isolated three-leg ladder, we expect that at $T \ll J$ 
the exchange interaction along a rung strongly couples the three $S=\frac{1}{2}$ spins 
into an effective $S_{eff}=\frac{1}{2}$, forming a $S_{eff}=\frac{1}{2}$ 
chain. 
Therefore, an isolated three-leg ladder would exhibit
$1/T_{2G} \propto 1/\sqrt{T}$ and $1/T_{1}\sim$ 
constant at low temperatures \cite{Starykh} as 
observed for 1d spin-chain Sr$_{2}$CuO$_{3}$\cite{Takigawa1D}.  This clearly 
contradicts with the exponential divergence we find.  
Three-dimensional spin freezing observed at $T_{SF}=$ 52 K 
\cite{muSR} is unlikely to be the 
origin of the observed divergence, either, because the onset of the 
exponential divergence ($\sim$ 300 K) is nearly a factor of 6 higher than 
$T_{SF}=$ 52 K.   Furthermore, the temperature 
dependence is exponential rather than the ordinary power 
law divergence of $1/T_{1}$ and $1/T_{2G}$ expected near 3d orderings.  The 
lack of 3d character 
is consistent with the fact that the exchange coupling along the b 
axis is frustrated due to opposing pairs of 90$^{o}$ Cu-O-Cu bonds, 
which suppresses 3d correlations \cite{gopalan}.   

The key point to note is that, along the c axis,
 the three-leg ladders are stacked directly on top of one another 
as shown in Fig.1(b).  We recall that the so-called infinite layer 
compound Ca$_{0.85}$Sr$_{0.15}$CuO$_{2}$ has 2d square-lattice layers with a 
similar c-axis stacking, and has an equivalent structure to Sr$_{2}$Cu$_{3}$O$_{5}$ except for 
the line defect between adjacent three-leg ladders.   
In Ca$_{0.85}$Sr$_{0.15}$CuO$_{2}$, N\'eel ordering driven by the 
large c axis coupling, $J_{c}$, 
occurs at extremely high temperature, 539 K, and 
the dimensional crossover from isolated 2d square-lattice behavior to 3d 
behavior occurs as high as 600 K \cite{Lombardi}.  This suggests that 
the three-leg ladders in Sr$_{2}$Cu$_{3}$O$_{5}$ will also couple 
strongly along the c-axis.  At low temperatures, 
the stacked three-leg ladders form a 2d plane of 
$S_{eff}=\frac{1}{2}$ with anisotropic 
exchange interactions, $J$ along the a axis and the effective c axis coupling 
for the 2d model, $J_{c}^{eff} \approx 3 J_{c}$, as shown in figure 1(c).  

Our viewpoint is supported 
by two sets of recent Monte Carlo simulations.  First, Greven and 
Birgeneau showed that inter-layer 
coupling $J_{c}$ along c-axis was essential to understand the 3d long-range 
order of Zn-doped two-leg ladder SrCu$_{2}$O$_{3}$ 
\cite{Greven2,Azuma2}.  
Second, more recent Monte-Carlo simulations by 
Y.J. Kim et al. showed that inter-ladder coupling two-orders 
of magnitude smaller than $J$ is sufficient to induce dimensional 
crossover from quasi-1d behavior of an isolated three-leg ladder to anisotropic 2d 
behavior of coupled three-leg ladders \cite{Young-jun}.  
In the present case, the inter-ladder coupling along c-axis is large, 
$J_{c}^{eff}/J \cong (0.15 - 0.22)$ using estimates $J_{c} \sim 
75 - 110$ K 
\cite{neutron,Greven2}, and $J = 1500$ K.  

Theoretically, the anisotropy $\alpha = J_{c}^{eff}/J$ of 
exchange interaction introduces anisotropy in the spin wave 
velocity $c_{0}$ and spin-spin correlation length $\xi$ for two 
orthogonal directions, and reduces the isotropic spin stiffness 
$2\pi\rho_{s}$ to an effective $2\pi\rho_{s}^{eff}$ \cite{anisotropic}.  That is,
$c_{\|}(\alpha) = \sqrt{(1 + \alpha)/2} \cdot c_{0},
c_{\perp}(\alpha) = \sqrt{\alpha} c_{\|}(\alpha)$, and 
$2\pi\rho_{s}^{eff} = \left( 1 - g_{0}(1)/ g_{c}(\alpha) \right) \sqrt{\alpha} 
2\pi\rho_{s}$, where
$g_{c}(\alpha)$ is the critical coupling constant, and $g_{0}(1)$ is the 
bare coupling constant for $\alpha=1$ \cite{anisotropic}.  Otherwise the theoretical 
framework of the renormalized classical regime in isotropic 2d square 
lattice \cite{Chak-Orb,Imai2D}, which was successfully employed to 
analyze $^{63}$Cu NQR/NMR relaxation rates in 
2d square-lattice La$_{2}$CuO$_{4}$ \cite{ImaiT1,Imai2D} and 
Sr$_{2}$CuO$_{2}$Cl$_{2}$ \cite{Thurber}, is applicable to the anisotropic case.  
By fitting $1/T_{1}T^{3/2}$ and $1/T_{2G}T$ in Fig.3 to exponential 
form, we obtain 
$2\pi\rho_{s}^{eff} = 290 \pm 30$ K.  This implies an anisotropy, $\alpha = 
0.16 (0.17) \pm 0.02$ for $J = 
1500 (1300) K$, hence $J_{c}^{eff}= 230 \pm 30$ K.  The obtained value of 
$\alpha = J_{c}^{eff}/J = 0.16 \pm 0.02$ is consistent with the estimate 
of 0.15 - 0.22 in the previous paragraph.

We can test the consistency of the preceding renormalized classical 
analysis by estimating $\alpha$ 
based on a different method without knowing $J$.  In the low-temperature renormalized classical limit, we expect the 
ratio $R(\alpha, 2\pi\rho_{s}^{eff}) = (T_{1}T^{3/2})/(T_{2G}T) \cdot (F_{ab}(\pi)/F_{c}(\pi))^{2} \propto 
\sqrt{(\hbar^{2} c_{\|}(\alpha)c_{\perp}(\alpha)/2\pi\rho_{s}^{eff})} \cdot 
(F_{ab}(\pi)/F_{c}(\pi))^{2}$, to be independent of 
temperature.  The ratio of hyperfine form factors, 
$F_{ab}(\pi)/F_{c}(\pi)$, can be determined experimentally as $0.42 
\pm 0.02$ from $T_{1c}/T_{1ab} = 3.4 \pm 0.2$, and $2\pi\rho_{s}^{eff} = 
290 \pm 30$ K from the fit in Figure 3.  This leaves $\alpha$ as the 
only unknown parameter in $R(\alpha, 2\pi\rho_{s}^{eff})$.  Shown in 
the inset to Fig.3, the 
ratio $R(\alpha, 2\pi\rho_{s}^{eff})$ indeed approaches a constant 
$61 \pm 5$ at low temperatures, which implies $\alpha = 0.15 \pm 0.03$, in agreement with 
our earlier 
estimate, $0.16 \pm 0.02$ \cite{Knight}.  

To conclude, we demonstrated that the temperature dependence of 
the spin-spin correlation length $\xi$ in Sr$_{2}$Cu$_{3}$O$_{5}$ is consistent with 
the isolated three-leg 
ladder behavior $\xi\sim\frac{1}{T}$ from 300 K to 725 K ($\sim J/2$).  
Below 300 K, we discovered 
dimensional crossover to anisotropic 2d regime, where spin 
correlations diverge exponentially.  Our result is the first 
experimental demonstration of the 
validity of the anisotropic non-linear sigma model, which was 
recently 
proposed for the stripe phase of high $T_{c}$ cuprates, in a model 
$S=\frac{1}{2}$ quantum Heisenberg antiferromagnet.  
This should encourage further theoretical analysis of the 
stripe physics of high $T_{c}$ cuprates based on the anisotropic non-linear sigma model.  

We are indebted to S. Sachdev, R. J. Birgeneau, and Y. J. 
Kim for their helpful 
discussions.  This 
work was supported by NSF DMR 99-71264, NSF DMR 98-08941, and in part by 
NSF 96-23858, the A.P. Sloan and the Mitsui foundations. 

% now the references. delete or change fake bibitem. delete next three
%   lines and directly read in your .bbl file if you use bibtex.

% \begin{figure}
% \caption{}
% \label{}
% \end{figure}

\begin{figure}
\caption{Structure of three leg ladder material, 
Sr$_{2}$Cu$_{3}$O$_{5}$, (a) top view, (b) side view, and (c) 
effective structure in anisotropic 2d regime below 300 K.
}
\label{structurefig}
\end{figure}

\begin{figure}
\caption{(a) $1/T_{1c}$ and (b) ($1/T_{2G}$)$_{NQR}$ for the 3-leg ladder 
Sr$_{2}$Cu$_{3}$O$_{5}$ (edge chain Cu site)[$\bullet$], in comparison 
to two-leg ladders SrCu$_{2}$O$_{3}$ [$\blacktriangledown$] and 
La$_{6}$Ca$_{8}$Cu$_{24}$O$_{41}$ [$\times$], 
2d square-lattice Sr$_{2}$CuO$_{2}$Cl$_{2}$ [$\bigcirc$], and 1d spin 
chain Sr$_{2}$CuO$_{3}$ [$\vartriangle$].  Inset to Fig.2(b): $T_{2G}$ at edge chain Cu site in 
Sr$_{2}$Cu$_{3}$O$_{5}$ and SrCu$_{2}$O$_{3}$.  Solid line shows fit above 300 K to 
predicted power law, $T_{2G} \propto T^{3/2}$ in the 
quasi-1d regime of an isolated three-leg ladder.
}
\label{T1T2gfig}
\end{figure}

\begin{figure}
\caption{$1/( T_{1c} T^{3/2} )$ (circles) and $1/( T_{2G} 
T )$ (triangles) 
versus $1/T$ for three-leg Sr$_{2}$Cu$_{3}$O$_{5}$ (solid symbols) and 2-d 
Sr$_{2}$CuO$_{2}$Cl$_{2}$ (open symbols).  The fit to renormalized 
classical (exponential) behavior for Sr$_{2}$Cu$_{3}$O$_{5}$ $T < 225$ K gives 
$2\pi\rho_{S}^{eff} = 290 \pm 30$ K, implying anisotropy $\alpha = 
0.16 \pm 0.02$ 
for $J=1500$ K.
Inset:  Ratio $R(\alpha, 2\pi\rho_{s}^{eff})=(T_{1c} T^{3/2} / T_{2G} T ) \cdot (F_{ab}^{2} / F_{c}^{2})$ for 
three-leg Sr$_{2}$Cu$_{3}$O$_{5}$.  The ratio should be 
constant in the low temperature limit deep inside the renormalized classical 
regime.  The solid line shows the value of $R(\alpha, 2\pi\rho_{s}^{eff})$ calculated for 
Sr$_{2}$Cu$_{3}$O$_{5}$ ($\alpha = 0.15$, $2\pi\rho_{S}^{eff} = 290$ 
K).}
\label{anisotropicfig}
\end{figure}

% \begin{table}
% \caption{}
% \label{}
% \end{tabular}
% \end{table}


\begin{references}
\bibitem{T1NQRNMR}The nuclear quadrupole interaction tensor $\nu_{Q\alpha}$ 
at the copper sites in ladders is not 
axially symmetric with respect to the c axis, $\eta = (\nu_{Qa} - 
\nu_{Qb})/\nu_{Qc}$=0.64 and 0.17 for the edge and central chain of 
ladders, respectively \cite{Ishida}.  It is straightforward to show 
that, based on 
the standard Redfield approach, $1/T_{1}$ measured by NQR is a combination 
of $1/T_{1}$ measured 
by high field NMR for different magnetic field directions, 
\begin{equation}
(\frac{1}{T_{1}})_{NQR} =( \frac{1}{9 + 3\eta^{2}})
[ 4\eta^{2} \frac{1}{T_{1ab}} + (9 - \eta^{2}) \frac{1}{T_{1c}}]
\end{equation}
where subscript $ab$ and $c$ denote the direction of magnetic field.
For the edge chain site, we measured $1/T_{1}$ with both NQR and high 
field NMR, and confirmed the consistency of the experimental ratios, 
$T_{1c}/T_{1ab} = 3.4 \pm 0.2$ and $T_{1c}/T_{1,NQR} 
= 0.70 \pm 0.03$ with the theoretical 
expression, $T_{1c}/T_{1,NQR} = 0.72 \pm 0.02$.  
For the purpose of systematic comparison with other materials, we 
multiplied a factor 0.70 to the NQR results to deduce $\frac{1}{T_{1c}}$.
\end{references}
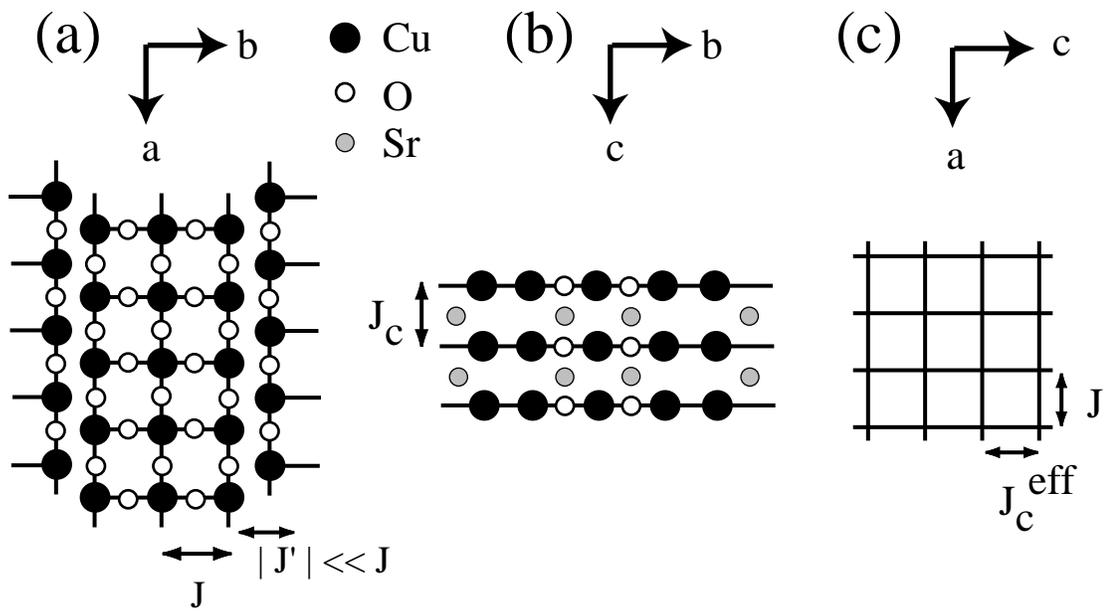
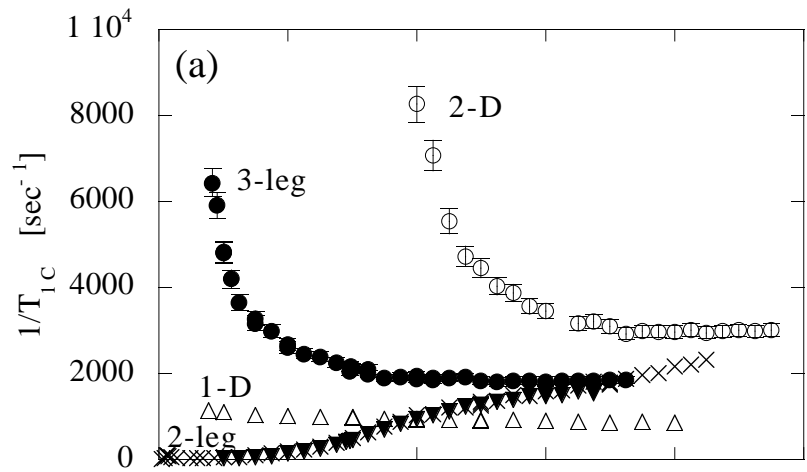
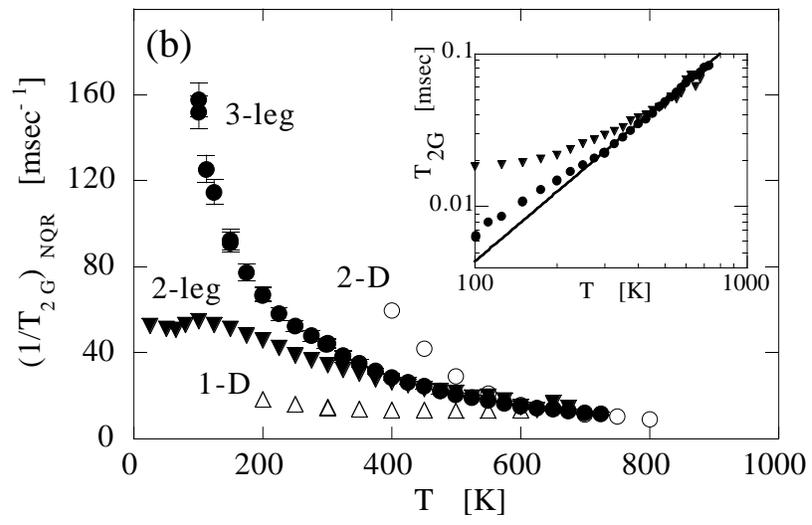
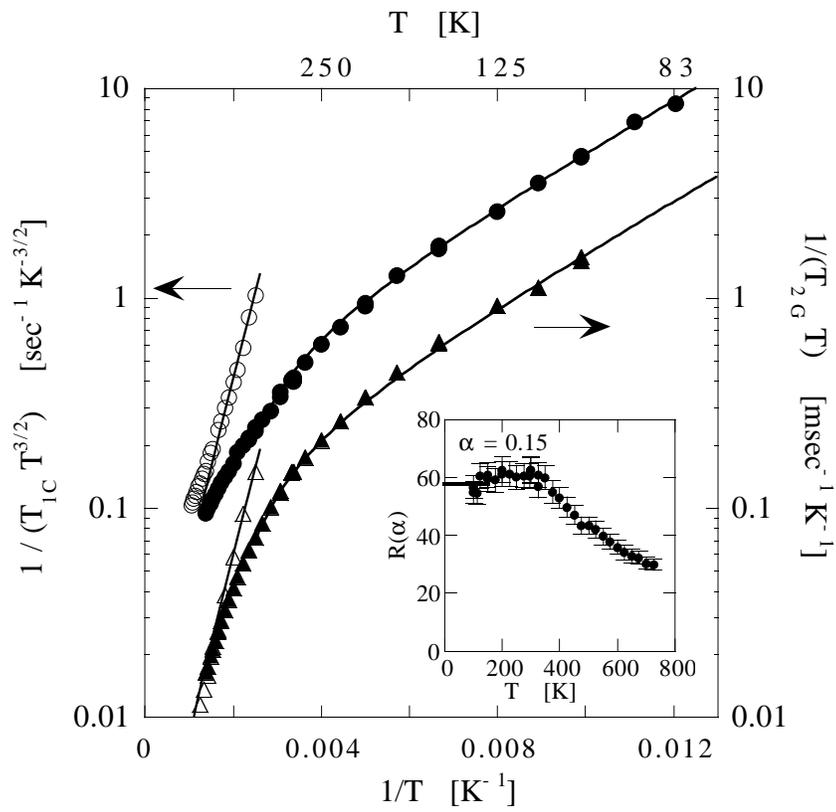
\end{document}